# ζ Pup: the merger of at least two massive stars?

Dany Vanbeveren

Astrophysical Institute, Vrije Universiteit Brussel
and
Leuven Engineering College, GroupT, Association KU Leuven

**Abstract.** We first discuss the stellar and wind parameters of ζ Pup resulting from detailed UV diagnostics. These parameters together with the runaway nature of the star can most easily be explained by dynamical binary-binary or binary-single star interactions in dense stellar clusters. In this case ζ Pup is most likely the merger of at least two massive stars.

## 1. Introduction

Massive binaries are hot again in the massive star community. This has not always been the case. In 1988, Tony was one of the organizers of the Val Morin (Quebec) meeting on massive stars in general, luminous blue variables in particular, and a majority of those who survived the rafting experience, returned home with the idea that massive binaries are rather unimportant curiosa. A minority who defended the role of massive binaries at that meeting risked to receive a virtual prize during the closing diner for beating a dead horse. However, some massive binary fans continued doing their job in Don Quichot style, and many interesting massive binary papers were published in this period. To illustrate, the statements *'many O-type single stars may have had a binary history'* or *'most of the Type Ib/c supernovae result from binary components with an initial mass between 10 $M_\odot$ and 20 $M_\odot$'* that popped up during the present meeting were illustrated in several papers published in the period before the year 2004, even in the previous millenium. For all those who want to know more what was known about massive binaries before 2004, the review of De Donder and Vanbeveren (2004, be prepared for a 115 page paper) gives the state of art.

The process where two binary components merge has been studied in the past mainly for those binaries where at least one star is a compact star. However, the interplay between cluster dynamics and binaries has shown that binaries with two normal hydrogen burning stars may merge as well, and that this process is very common in dense clusters. The present paper is a summary of the paper of Pauldrach et al. (2011) where we show that one of the brightest stars in the sky, ζ Pup, may be the merger of at least two massive core hydrogen burning stars.

## 2. The stellar and wind parameters of ζ Pup based on UV diagnostics

The Munich code to calculate detailed atmospheric models for hot stars assumes homogeneous, stationary, and spherically symmetric radiation driven winds (Pauldrach et al., 2001). It accounts for the following physical processes:

- The hydrodynamic equations are solved where the crucial term is the radiative acceleration. Note that the wind velocity is computed self-consistently and no velocity law is adopted a priori

- The occupation numbers are determined by the rate equations

- The spherical transfer equation is solved (including EUV line blocking). Stark-broadening has been included in the procedure. Moreover, the shock source functions produced by radiative cooling zones which originate from a simulation of shock heated matter, together with K-shell absorption, are also included.

- The temperature structure is determined by the microscopic energy equation where the effects of line blanketing are taken into account.

The iterative solution of the total system of coupled equations yields the hydrodynamic structure of the wind together with the synthetic spectra and ionizing fluxes.
The question whether $\dot{M}$, $v_{inf}$ and the spectral energy distribution predicted by these models are already realistic enough to be used for diagnostic issues requires of course an ultimate test. This has been done for ζ Pup, one of the most luminous O supergiants in the Galaxy. The winds of hot stars are driven by photon momentum transfer through line absorption, and hence the wind momentum rate (the mass loss rate and the terminal velocity) must be a function of the abundances; and this behavior influences the signatures of the spectra in the same way as the abundances do that directly. Our simulations for ζ Pup clearly indicate that it is by no means correct to simply assume solar abundances, in other words, the stellar parameters must be determined along with and simultaneously with the individual values of the abundances. The iteration procedure therefore begins with reasonable estimates of $T_{eff}$, the stellar radius R, the stellar mass M and the abundances. The model atmosphere is solved and the velocity field, the mass loss rate and the synthetic spectrum are calculated. The parameters are then adjusted and the process is repeated until a good fit to all features in the observed UV spectrum is obtained. This procedure results in the stellar and wind parameters for ζ Pup given in Table 1. The very good correspondence between the observed UV spectrum and the model prediction is illustrated in Figure 1.

The following bullets are important for the scope of the present paper:

- ζ Pup is a rapid rotator and significantly enriched in He; note that the latter is the case for many supergiants

- The mass loss rate is very large (~$1.410^{-5}$ $M_{\odot}$/yr)

- The chemical abundances deviate considerably from the solar values.

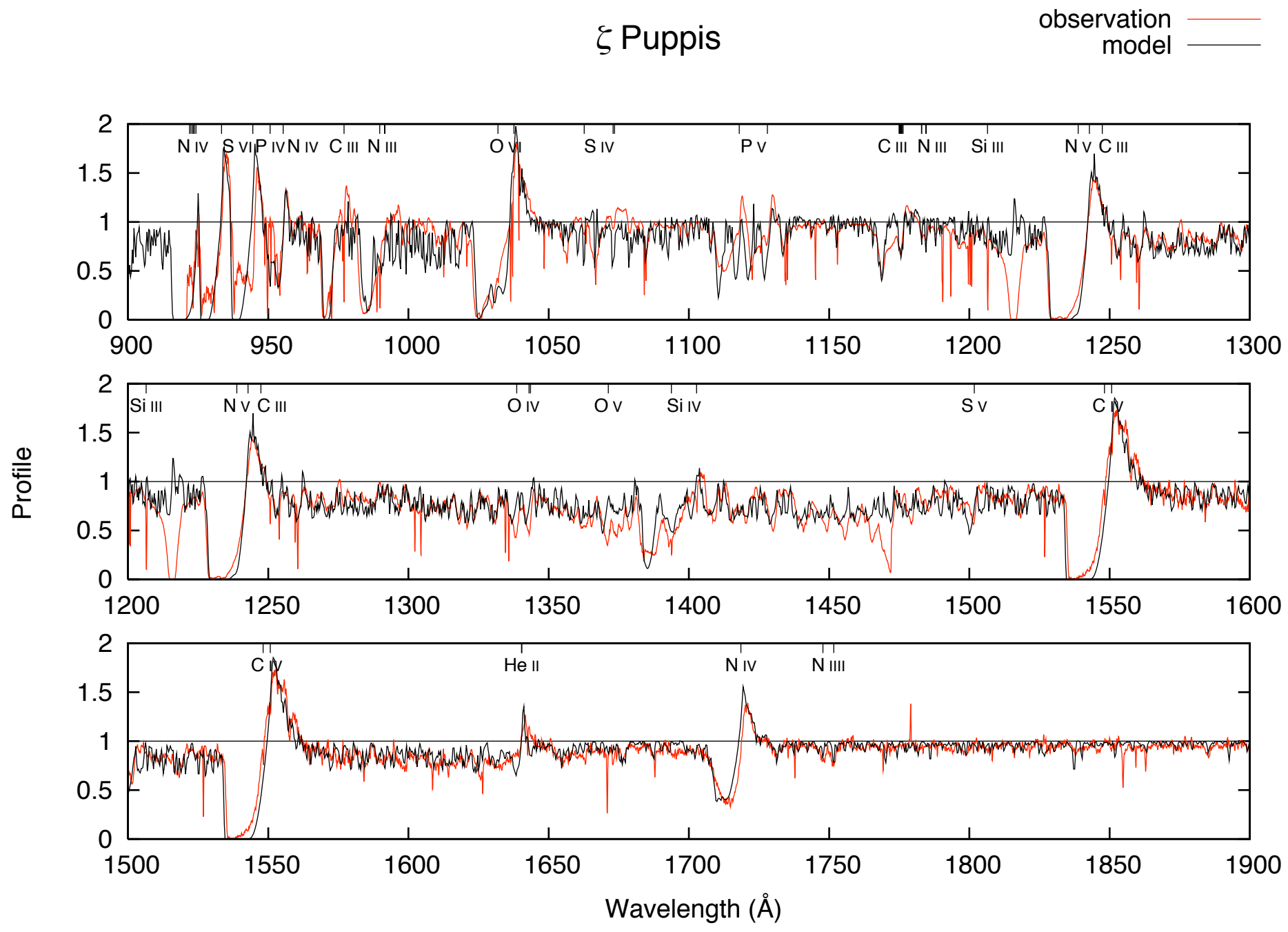

Figure 1 The theoretically predicted UV spectrum of ζ Pup (using the parameters given in Table 1) compared to the observed one.

Table 1. The parameters of ζ Pup (the abundances are in Solar units)

| Parameter | Value | Parameter | Value |
|---|---|---|---|
| $T_{eff}$ (K) | 40000 | $R/R_{\odot}$ | 28.0 |
| Log $L/L_{\odot}$ | 6.26 | $M/M_{\odot}$ | 71.9 |
| $\dot{M}$ ($10^{-6}$ $M_{\odot}$/yr) | 13.8 | $v_{rot}$ (km/s) | 220 |
| He | 1.6 | C | 2.7 |
| N | 8.3 | O | 0.15 |
| Fe | 2.2 | Si | 8.8 |

## 3. ζ Pup runs away from the young Vela R2 association

Since the stellar radius has been derived from the UV diagnostics, its value results directly in the distance of the star, which is 692 pc. This is considerably larger than the Hipparcos distance but one may wonder whether or not Hipparcos distances larger than 200pc are reliable. Interestingly, with this distance the past path of ζ Pup goes through the association Vela R2 (Van Rensbergen et al., 1996). The association is at most a few Myr old (compatible with the age of ζ Pup) and the past path indicates that it left Vela R2 some 1.5 Myr ago. Its distance combined with the radial velocity and the proper motion results in a space velocity for the supergiant > 70 km/s. Accounting for its very large mass (Table 1), ζ Pup is a hell of a runaway.

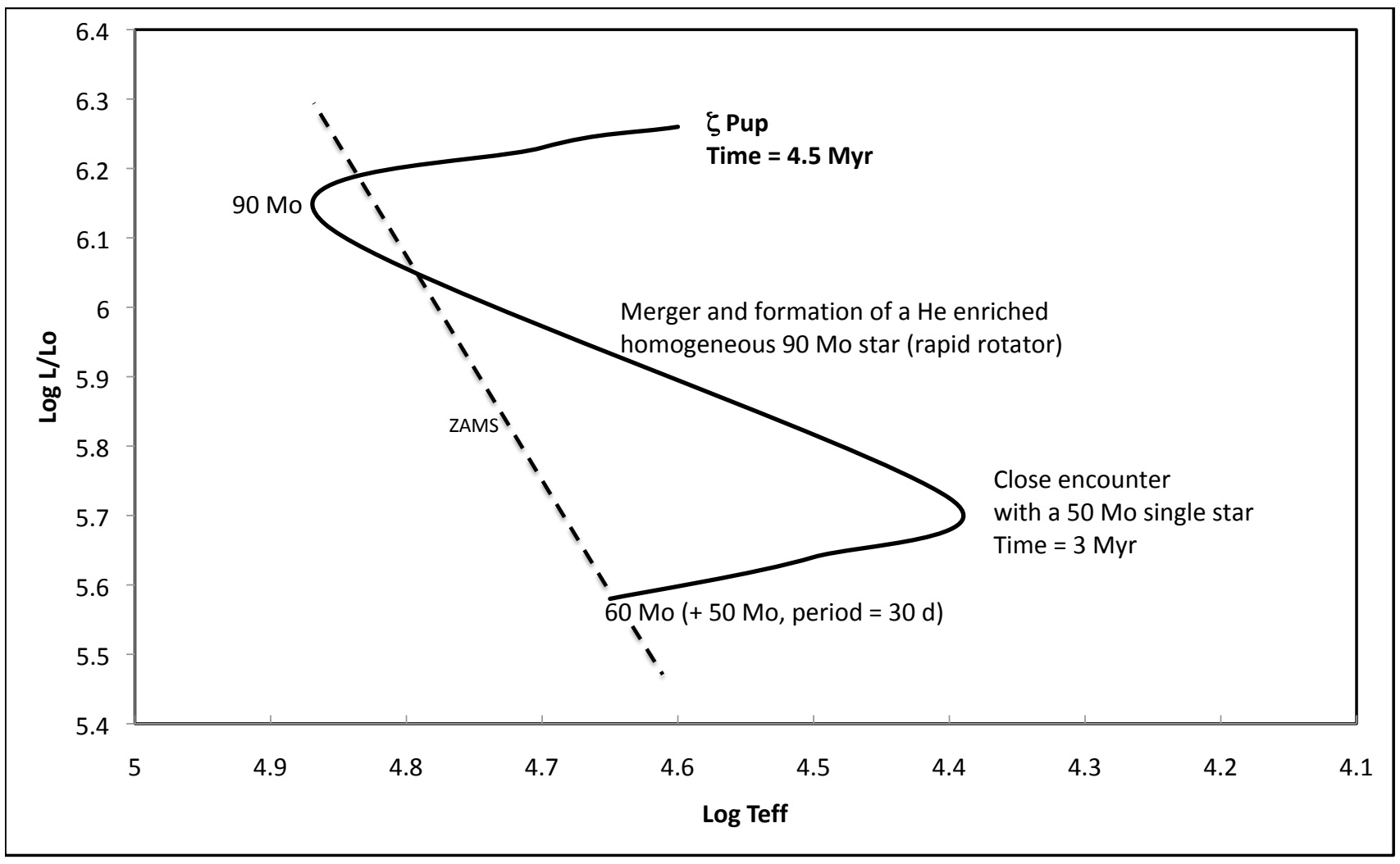

Figure 2 A most plausible evolutionary model for the progenitor of ζ Pup.

## 4. ζ Pup: the merger of at least two massive stars

If the mass and the distance (thus the stellar radius) of ζ Pup discussed above are correct then it requires exotic assumptions in order to explain this runaway with the classical binary scenario where the runaway status of a star is due to the supernova explosion of its companion. The ejection of stars due to dynamical interactions of stars and binaries in dense clusters offers an interesting alternative. Vanbeveren et al. (2009) discussed such a scenario for ζ Pup. Here we extend the discussion in view of the stellar parameters given in Table 1. We performed about 1 million binary-single or binary-binary scattering experiments with different binary and single star masses, with binary periods ranging between 5-1000 days. The binary eccentricities were taken from a thermal distribution. Note that many combinations of the parameters produce the required mass and runaway velocity but a most probable one involves at least one massive binary with a period = 30 days. Figure 2 then illustrates a very plausible evolutionary scenario for ζ Pup, a scenario that also reproduces the stellar parameters discussed in the present paper. A 60+50 $M_{\odot}$ binary with a period = 30 days formed initially in the Vela R2 cluster sinks to the cluster center due to mass segregation. After about 3 Myr, when both stars have transformed already a significant fraction of hydrogen in helium in the core, the binary encounters a 50 $M_{\odot}$ massive single star (the latter also sunk to the cluster center due to mass segregation). As a consequence of the dynamical interaction, the two binary components merge and the merger acquires a runaway velocity 60-80 km/s and will leave Vela R2. As shown by Suzuki et al. (2007) the merger of two massive stars results in a new star with a mass approximately equal to the sum of masses of the two stars (during the merger process some 10 $M_{\odot}$ may be lost) and the merger is efficiently mixed (the merger is largely homogenized). Since before the encounter, the two cores of the binary components were already significantly He enriched, the mixing results in a homogeneous star with large He abundance (a large He abundance is observed in the case of ζ Pup). The merger process of two binary components obviously results in a rapid rotator, and the observations reveal that ζ Pup is a rapid rotator. About 1.5 Myr after the merger process, the star has all the properties to be ζ Pup.